\documentclass[journal=ancac3,manuscript=article]{achemso}

\usepackage[T1]{fontenc} % Use modern font encodings
\usepackage{amsfonts}
\usepackage{amssymb}
\usepackage{amsmath}
\usepackage{amsthm}
\usepackage{graphicx}
\usepackage[breaklinks=true,colorlinks=true,linkcolor=blue,urlcolor=blue,citecolor=blue]{hyperref}
\usepackage{color}
\usepackage{soul}
\usepackage[separate-uncertainty,multi-part-units = single]{siunitx}
\usepackage{gensymb}

\author{Matthew Hamer}
\affiliation{School of Physics and Astronomy, University of Manchester, Oxford Road, Manchester, M13 9PL, UK}
\alsoaffiliation{National Graphene Institute, University of Manchester, Oxford Road, Manchester, M13 9PL, UK}
\author{Johanna Zultak}
\affiliation{School of Physics and Astronomy, University of Manchester, Oxford Road, Manchester, M13 9PL, UK}
\alsoaffiliation{National Graphene Institute, University of Manchester, Oxford Road, Manchester, M13 9PL, UK}
\author{Anastasia V. Tyurnina}
\affiliation{School of Physics and Astronomy, University of Manchester, Oxford Road, Manchester, M13 9PL, UK}
\alsoaffiliation{National Graphene Institute, University of Manchester, Oxford Road, Manchester, M13 9PL, UK}
\author{Viktor Z\'{o}lyomi}
\affiliation{School of Physics and Astronomy, University of Manchester, Oxford Road, Manchester, M13 9PL, UK}
\alsoaffiliation{National Graphene Institute, University of Manchester, Oxford Road, Manchester, M13 9PL, UK}
\author{Daniel Terry}
\affiliation{School of Physics and Astronomy, University of Manchester, Oxford Road, Manchester, M13 9PL, UK}
\alsoaffiliation{National Graphene Institute, University of Manchester, Oxford Road, Manchester, M13 9PL, UK}
\author{Alexei Barinov}
\affiliation{Elettra - Sincrotrone Trieste, S.C.p.A., Basovizza (TS), 34149, Italy}
\author{Alistair Garner}
\author{Jack Donoghue}
\author{Aidan P. Rooney}
\affiliation{School of Materials, University of Manchester, Oxford Road, Manchester, M13 9PL, UK}
\author{Viktor Kandyba}
\author{Alessio Giampietri}
\affiliation{Elettra - Sincrotrone Trieste, S.C.p.A., Basovizza (TS), 34149, Italy}
\author{Abigail J. Graham}
\author{Natalie C. Teutsch}
\author{Xue Xia}
\affiliation{Department of Physics, University of Warwick, Coventry, CV4 7AL, UK}
\author{Maciej Koperski}
\affiliation{School of Physics and Astronomy, University of Manchester, Oxford Road, Manchester, M13 9PL, UK}
\alsoaffiliation{National Graphene Institute, University of Manchester, Oxford Road, Manchester, M13 9PL, UK}
\author{Sarah J. Haigh}
\affiliation{School of Materials, University of Manchester, Oxford Road, Manchester, M13 9PL, UK}
\alsoaffiliation{National Graphene Institute, University of Manchester, Oxford Road, Manchester, M13 9PL, UK}
\author{Vladimir I. Fal'ko}
\email{*Vladimir.Falko@manchester.ac.uk}
\affiliation{School of Physics and Astronomy, University of Manchester, Oxford Road, Manchester, M13 9PL, UK}
\alsoaffiliation{National Graphene Institute, University of Manchester, Oxford Road, Manchester, M13 9PL, UK}
\author{Roman Gorbachev}
\email{*Roman@manchester.ac.uk}
\affiliation{School of Physics and Astronomy, University of Manchester, Oxford Road, Manchester, M13 9PL, UK}
\alsoaffiliation{National Graphene Institute, University of Manchester, Oxford Road, Manchester, M13 9PL, UK}
\author{Neil R. Wilson}
\email{*Neil.Wilson@warwick.ac.uk}
\affiliation{Department of Physics, University of Warwick, Coventry, CV4 7AL, UK}

\title[InSe electronic structure]{Indirect to direct gap crossover in two-dimensional InSe revealed by ARPES}
\keywords{ARPES, indium selenide, 2D materials, density functional theory, photoluminescence, spin-orbit coupling}

\begin{document}

%\begin{tocentry}

%\includegraphics[height=3.5cm]{figTOC}

%\end{tocentry}


\begin{abstract}

Atomically thin films of III-VI post-transition metal chalcogenides (InSe and GaSe) form an interesting class of two-dimensional semiconductor that feature strong variations of their band gap as a function of the number of layers in the crystal \cite{Mudd2013,Mudd2015,Bandurin2016,Hamer2018} and, specifically for InSe, an earlier predicted crossover from a direct gap in the bulk \cite{Larsen1977, Amokrane1999} to a weakly indirect band gap in monolayers and bilayers \cite{Zolyomi2013,Zolyomi2014,Wu2014,Cao2015,Magorrian2016}.
Here, we apply angle resolved photoemission spectroscopy with submicrometer spatial resolution ($\mu$ARPES) to visualise the layer-dependent valence band structure of mechanically exfoliated crystals of InSe. We show that for 1 layer and 2 layer InSe the valence band maxima are away from the $\mathbf{\Gamma}$-point, forming an indirect gap, with the conduction band edge known to be at the $\mathbf{\Gamma}$-point. In contrast, for six or more layers the bandgap becomes direct, in good agreement with theoretical predictions. The high-quality monolayer and bilayer samples enables us to resolve, in the photoluminescence spectra, the band-edge exciton (A) from the exciton (B) involving holes in a pair of deeper valence bands, degenerate at $\mathbf{\Gamma}$, with the splitting that agrees with both $\mu$ARPES data and the results of DFT modelling.
Due to the difference in symmetry between these two valence bands,
light emitted by the A-exciton should be predominantly polarised perpendicular
to the plane of the two-dimensional crystal, which we have verified for few-layer InSe crystals.

\end{abstract}

\bigskip
Two-dimensional materials (2DM) and their van der Waals heterostructures, constructed by the mechanical assembly of individual 2D crystals, have a great potential for optoelectronic applications \cite{Novoselov2016}. The fast growing family of 2DM \cite{Geim2013} includes 2D insulators, 2D semiconductors with various band gaps, 2D metals and even 2D superconductors, with electronic and optical properties that often differ from their bulk allotropes \cite{Schaibley2016}. In this family, post-transition metal monochalcogenides (PTMC), III-VI compounds such as GaSe and InSe, are emerging as important materials to study, due to their interesting layer-dependent optical properties and exceptionally high carrier mobility \cite{Mudd2013,Mudd2015,Bandurin2016,Lei2014,Zolyomi2014,Mudd2016,Hamer2018}.
Both GaSe and InSe display a pronounced quantum confinement effect: an increase of the band gap upon decreasing the number of layers, $L$, which is stronger in InSe \cite{Magorrian2016,Bandurin2016} than in GaSe \cite{Terry2018} films as revealed recently by photoluminescence (PL) spectroscopy of atomically thin films of these compounds.
The latter effect is partly due to the interlayer hybridisation of S-orbitals of metal atoms and P$_z$-orbitals of Se dominating among the states at the edge of the conduction band at the $\mathbf{\Gamma}$-point, where electrons also have a light in-plane mass whose $L$-dependent values, determined from the temperature variation of Shubnikov de Haas oscillations \cite{Bandurin2016}, coincide with the theoretically predicted masses \cite{Magorrian2016}.
Such a spectral evolution is accompanied by the theoretically predicted flattening of the top valence band dispersion, which has the potential to lead to a phase transition of
p-doped GaSe monolayer films into, e.g., a ferromagnetic state \cite{Wu2014,Cao2015}, but is more likely to localise holes, as observed by systematically high (M$\Omega$ range) resistances measured in p-doped few-layer InSe films.
Moreover, for both InSe and GaSe the valence band edge in monolayer and bilayer films was predicted to shift away from the $\mathbf{\Gamma}$-point, as recently confirmed for GaSe by angle resolved photoemission spectroscopy (ARPES) studies \cite{BenAziza2017,BenAziza2018}.

Here, we use ARPES to determine the valence band structure of monolayer and few-layer crystals of $\gamma$-InSe. Conventional ARPES is limited to large (typically \SI{> 100}{\um}) atomically flat samples, which for most atomically thin materials necessitates epitaxial growth on single crystal substrates.
For this reason, previous ARPES experiments on monolayer metal monochalcogenides have used materials grown by molecular beam epitaxy \cite{Kibirev2018,Chen2018,BenAziza2017,BenAziza2018}.
Most transport and optical investigations instead use mechanically exfoliated flakes which, though higher-quality, are typically only a few \SI{}{\um} across. We have recently demonstrated that sub-micrometer spatially resolved ARPES ($\mu$ARPES) enables high resolution measurements from mechanically exfoliated flakes \cite{Wilson2017}.
In this way we can directly determine the valence band electronic structure in the same samples as were used for electrical and optical studies in these 2DM. Here, we combine $\mu$ARPES with optical spectroscopy and \textit{ab initio} calculations to directly demonstrate the crossover from direct to indirect gap and gain insight into the peculiar layer-number-dependent electronic structure of atomic films of InSe.

\begin{figure}[t]
 \centering
 \includegraphics[width=8cm]{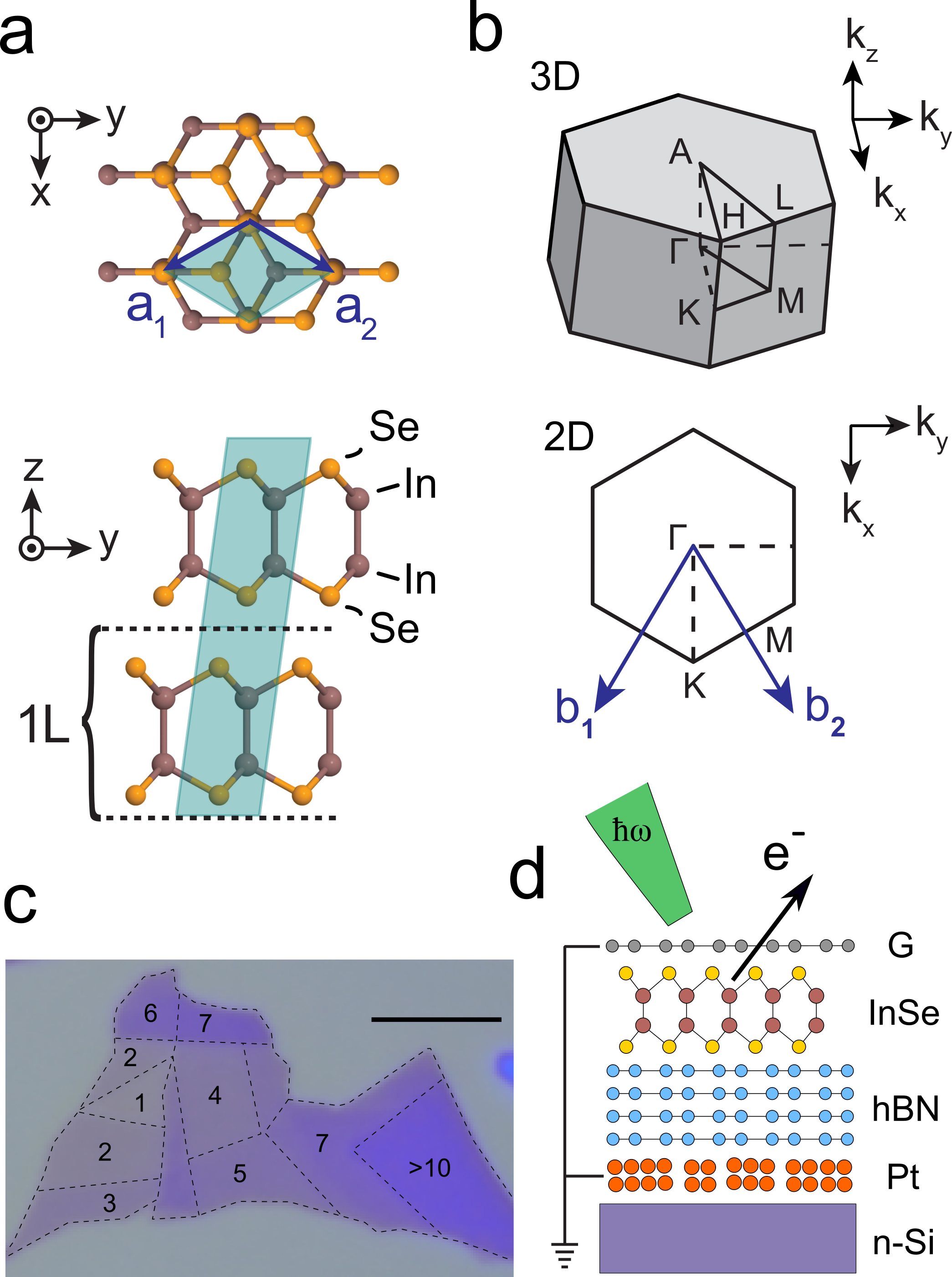}
 \caption{\label{schematic} \textbf{InSe sample structure.} Crystal structure, (a), and Brillouin zone, (b), of $\gamma$-InSe.
The blue shaded regions in (a) indicate the unit cell, lattice vectors are shown in blue.
(c) Schematic of the 2D heterostructure used for ARPES measurements.
(d) Optical microscope image of a mechanically exfoliated InSe flake, regions of different optical contrast
correspond to regions of different thickness (scale bar \SI{5}{\um}). }
\end{figure}

\subsection{Results and Discussion}

Layers of InSe have a honeycomb atomic arrangement produced by four planes of atoms, Se-In-In-Se.
Stacked in the most commonly observed $\gamma$-InSe polytype, Fig. \ref{schematic}a, few-layer-thick InSe has
a hexagonal 2D Brillouin zone, Fig. \ref{schematic}b.
An optical microscope image of a typical InSe crystal produced by mechanical exfoliation onto an oxidised silicon wafer
is shown in Fig. \ref{schematic}d, where individual $1$ to $7L$ terraces were identified by their optical contrast, confirmed by
atomic force microscopy. Individual regions within the flake were identified
by scanning photoemission microscopy (SPEM) \cite{Wilson2017} before performing detailed ARPES studies of each uniform
few \SI{}{\um} area. To acquire high-resolution spectra, the flakes must be transferred (see Methods)
onto an atomically flat substrate, electrically grounded to dissipate the photoemission current,
and a clean surface recovered \cite{Wilson2017}. To meet these criteria, mechanically exfoliated InSe flakes were sandwiched between hexagonal boron nitride or graphite underneath and an encapsulating layer of graphene on top, as shown schematically in Figure \ref{schematic}c. This provides charge dissipation and protects the material from decomposition during air exposure and subsequent vacuum annealing necessary to clean the surface. The stack was constructed in a purified Ar environment using a remotely controlled micromanipulation system \cite{Cao2015b} that enabled us to preserve the pristine crystalline structure by avoiding the oxidation of InSe \cite{Hamer2018}.

\begin{figure}[H]
 \centering
 \includegraphics[width=8cm]{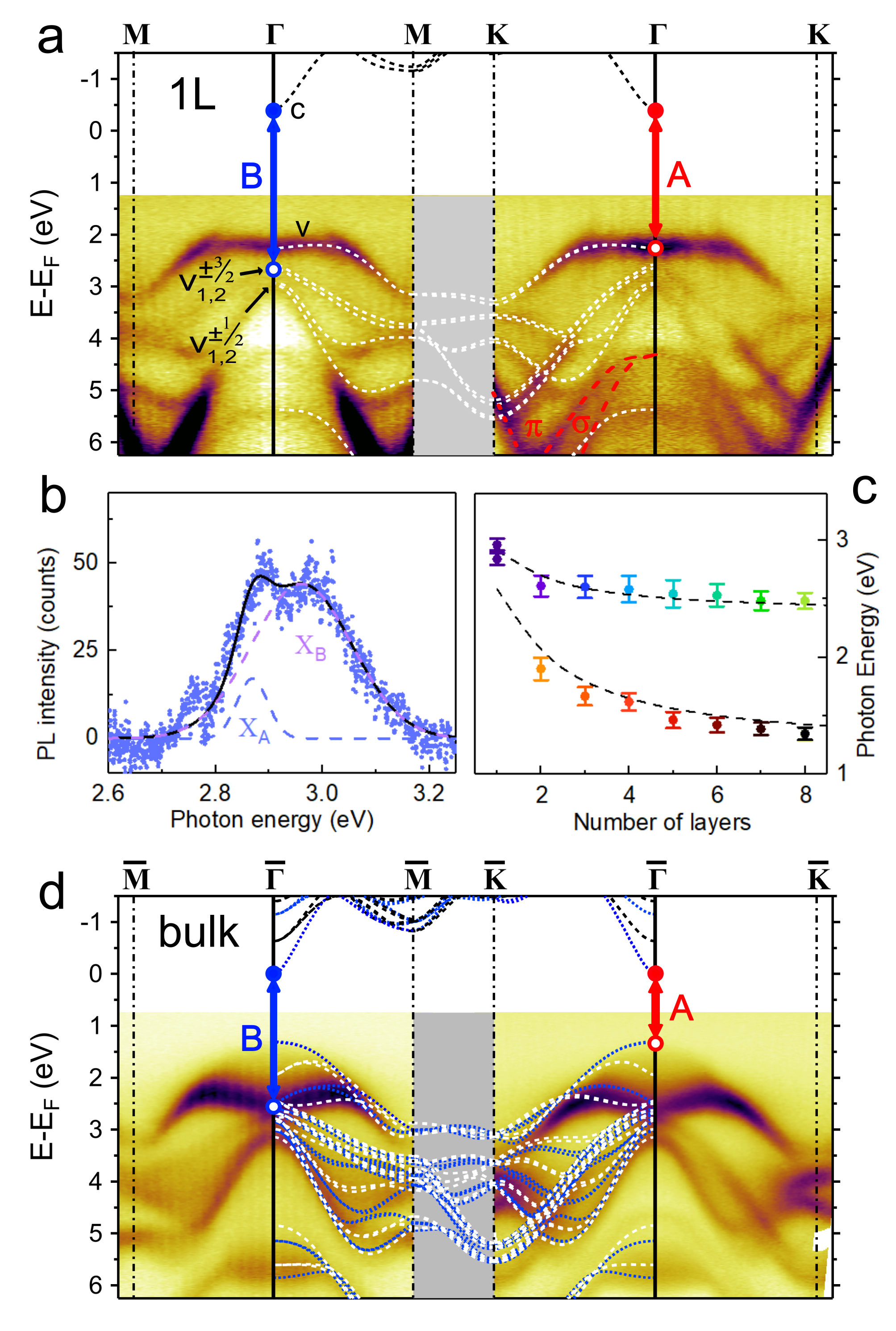}
 \caption{\label{Ek} \textbf{$\mathbf{\gamma}$-InSe valence band dispersion.} $\mu$ARPES measurement of the valence band structure of mechanically exfoliated 1L, (a), and bulk, (d), InSe along the high symmetry directions as labelled. DFT band structures are overlaid as white (black) dashed lines for the valence (conduction) band; in (d) the blue dotted lines correspond to the theoretical predictions for $k_z=0.5$ and the white / black to $k_z=0$ in units of out-of-plane reciprocal lattice constant. The red dashed lines are guides to the eye for the graphene bands as labelled. The experimental data has been reflected about $\mathbf{\Gamma}$ in each case to aid comparison with theory. Optical transitions corresponding to the A (B) exciton are labelled in red (blue). The photoluminescence spectrum of the monolayer is also shown, (b), along with the dependence of the optical transition energies in PL on the number of layers, (c).}
\end{figure}

Figure \ref{Ek} shows the results of $\mu$ARPES measurements of the valence band dispersion of monolayer and bulk InSe, both compared to the results of DFT modelling (white dashed lines). The energy-momentum $\mathrm{I}\!\left(E,\mathbf{k_\parallel}\right)$ spectra are plotted along the high-symmetry directions, as marked. 
In the monolayer data, Fig. \ref{Ek}a, graphene's $\pi$ and $\sigma$ bands are highlighted by red dashed lines.
For all samples measured, the Dirac point energy, $E_D$, of the graphene on top of the InSe was within \SI{50}{meV} of the Fermi energy, $E_F$, indicating little charge transfer between InSe and graphene. 
The binding energy of $v$ then gives the layer-dependent valence band offset between InSe and graphene, Table \ref{table2}. The large band offset for monolayer InSe confirms a significant Schottky barrier would be formed if graphene were used as a contact to InSe, explaining the high resistance previously reported  \cite{Bandurin2016} (however, as we discuss below, the band offset is smaller for few-layer crystals, in agreement with high quality tunable contacts observed in Ref. \cite{Hamer2018}).
The measured monolayer InSe bands appear to be in a good agreement with valence band spectra calculated using VASP code \cite{VASP} with experimentally measured lattice parameters, as overlaid in Figure \ref{Ek}a. The upper valence band, labelled $v$, is almost flat near $\Gamma$, dispersing to higher binding energy at the Brillouin zone boundaries. Our DFT calculations (see Methods) take into account spin-orbit coupling (SOC), which is necessary \cite{Magorrian2017,BenAziza2018} for the accurate description of two quadruplets of the lower-lying valence bands, $v_1$ and $v_2$, dominated by the P$_x$ and P$_y$ orbitals of Se atoms. These bands are distinguished by the $z \rightarrow -z$ mirror symmetry
of their wave functions, which are even in $v_1$ bands and odd in $v_2$, and each of these quadruplets is split
\cite{Magorrian2017} by atomic SOC into two doublets $v_{1,2}^{\pm \frac32}$ and $v_{1,2}^{\pm \frac12}$ with the
total (spin plus orbital) angular momentum projections
onto the z-axis of $\pm \frac32$ (higher energy) and $\pm \frac12$ (lower energy), respectively.
To further aid the discussion on optical properties, we also show the conduction band, $c$, as predicted by theory
with a scissor correction implemented \cite{Magorrian2016} to counter the underestimation of the gap in DFT (see Methods).

The ARPES spectra in Fig. \ref{Ek}a can be compared to photoluminescence (PL) observed in monolayer InSe, shown in Fig. \ref{Ek}b.
Here, the A-exciton corresponds to the transition from $c \leftrightarrow v$ at $\mathbf{\Gamma}$, as marked by the red arrow on Figure \ref{Ek}a,
and the B-exciton to the transition $c \leftrightarrow v_1^{\pm \frac32}$, as marked by the blue arrow.
Based on the PL spectrum shown in Fig. \ref{Ek}b, we are, now, able to resolve A- and B-exciton lines in monolayer InSe
(which was not possible in the earlier optical studies of this system \cite{Bandurin2016}) and compare those to the ARPES data in Table \ref{table1}.
From the ARPES spectra we determine the energy difference between $v$ and the upper doublets $v_{1,2}^{\pm \frac32}$ of $v_1$ and $v_2$ quadruplets
to be \SI{300\pm100}{meV} (the uncertainty is large due to the weak photoemitted intensity for $v_1$ and the broad linewidth of the intense $v$ band).
This appears to be larger than the A-B exciton splitting, \SI{120\pm70}{meV}, and the estimate, \SI{170}{meV},
of the earlier-developed tight-binding model (TBM) \cite{Magorrian2017}.
Such a difference can be attributed to the hybridisation of selenium's P$_z$ orbitals with carbon,
which pushes the $v$ band upward, increasing its distance from $v_{1,2}$ bands whose P$_{x,y}$ orbitals are immune to the interlayer hybridisation.

\begin{figure}[t]
 \centering
 \includegraphics[width=8cm]{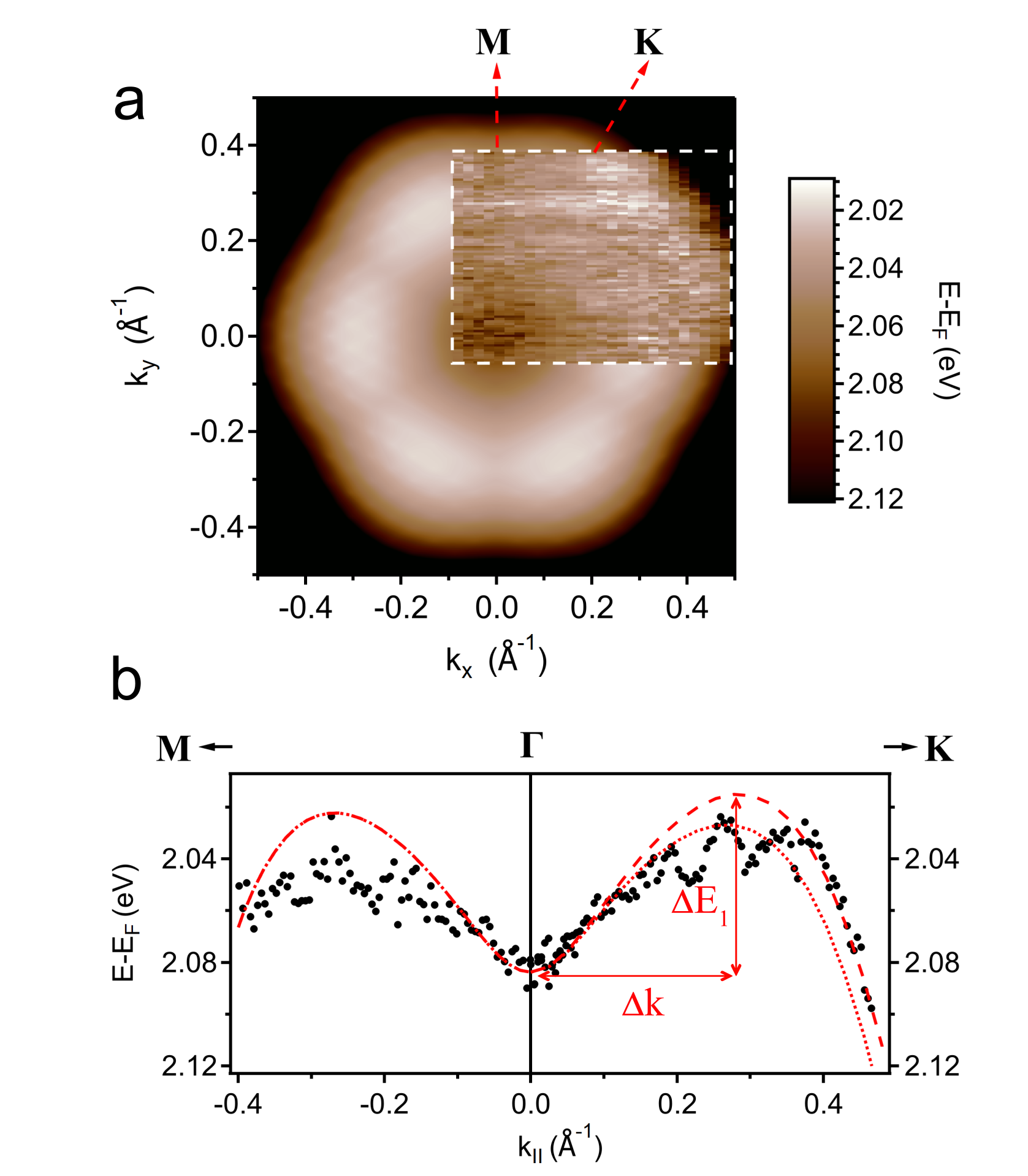}
 \caption{\label{mexican} \textbf{Valence band inversion in 1L InSe around $\mathbf{\Gamma}$.} (a) Upper valence band energy near $\mathbf{\Gamma}$ as predicted by DFT with experimental data overlaid within the white dashed rectangle, the color scale is identical for both and is given to the right. (b) Line profiles of the upper valence band dispersion close to $\mathbf{\Gamma}$ in the $\mathbf{M} \rightarrow \mathbf{\Gamma} \rightarrow \mathbf{K}$ directions as marked: the black circles are experimental data and the red dashed and dotted lines are DFT predictions for spin up and down bands.}
\end{figure}

For monolayer InSe the ARPES spectra reveal a band inversion around $\mathbf{\Gamma}$, as shown in detail in Figure \ref{mexican}.
The energy of the upper valence band, as predicted by DFT, is plotted as a function of $k_x$ and $k_y$ in the vicinity
of $\mathbf{\Gamma}$ in Figure \ref{mexican}a, where a lighter color corresponds to a lower binding energy.
There are six valence band maxima (VBM), each $\Delta k$ from the zone center in the $\mathbf{K}/\mathbf{K^\prime}$ directions,
with corresponding saddle points on the $\mathbf{\Gamma}-\mathbf{M}$ line. Within the white dashed box,
experimental band energies extracted from the ARPES spectra (see Methods) are overlaid, and 
the VBM and saddle points can be discerned despite the noise of the measurements.
Line profiles through both the experimental data and DFT simulations are shown in Fig. \ref{mexican}b,
allowing the position of the VBM and the depth of the band inversion to be determined as
$\Delta k$\SI{=0.3\pm0.1}{\per\angstrom} and $\Delta E_1$\SI{=50\pm20}{meV}  respectively.
In addition, we find that the saddle point is \SI{20 \pm 20}{meV} relative to the VBM.
These values agree within uncertainty with our DFT predictions: VBM at $\Delta k$\SI{=0.28}{\per\angstrom} from $\mathbf{\Gamma}$
and $\Delta E_1$\SI{=69}{meV}, with the saddle point in the $\mathbf{M}$ direction \SI{7}{meV} higher in binding energy than the VBM,
confirming that the monolayer is, indeed, an indirect band gap semiconductor.

\vspace{0.5cm}

\begin{table}
\begin{tabular}{c|c c}

\hline

InSe 1L & $v-v_1^{\pm \frac32}$ (meV) & $v_1^{\pm \frac32}-v_1^{\pm \frac12}$ (meV) \\

\hline

TBM \cite{Magorrian2017} &  $170$      & $380$   \\

DFT (this work)          &  $310$      & $360$   \\

ARPES                    &  $300\pm100$ & $400\pm100$  \\

PL, $X_B-X_A$            &  $120\pm70$  &  ---    \\

\hline

\end{tabular}

\caption{\label{table1} \textbf{InSe monolayer valence band parameters at $\Gamma$.}
Monolayer energy splittings between the upper valence band and the lower valence bands,
comparing the ARPES and PL data to previous\cite{Magorrian2017}
calculations and DFT modelling in the present work.}

\end{table}

\vspace{0.5cm}

Unlike the monolayer, bulk $\gamma$-InSe is clearly a direct band gap material, Fig. \ref{Ek}d,
in agreement with the existing literature \cite{Larsen1977, Amokrane1999, Politano2017}.
The ARPES spectra of a bulk crystal show broad features with multiple states dispersed in the $k_z$ direction,
perpendicular to the layers, since $\mu$ARPES detects photoemitted electrons with a wide (here, undetermined) range of $k_z$.
The white and blue dashed lines correspond to DFT band structure calculations with $k_z=0$ and $k_z=0.5$
(in units of out-of-plane reciprocal lattice parameter) respectively. Where these overlay, and are similar to the monolayer data,
the bands are typically 2D in nature with little dispersion in the $k_z$ direction.

\begin{figure}[t]
 \centering
 \includegraphics[width=12cm]{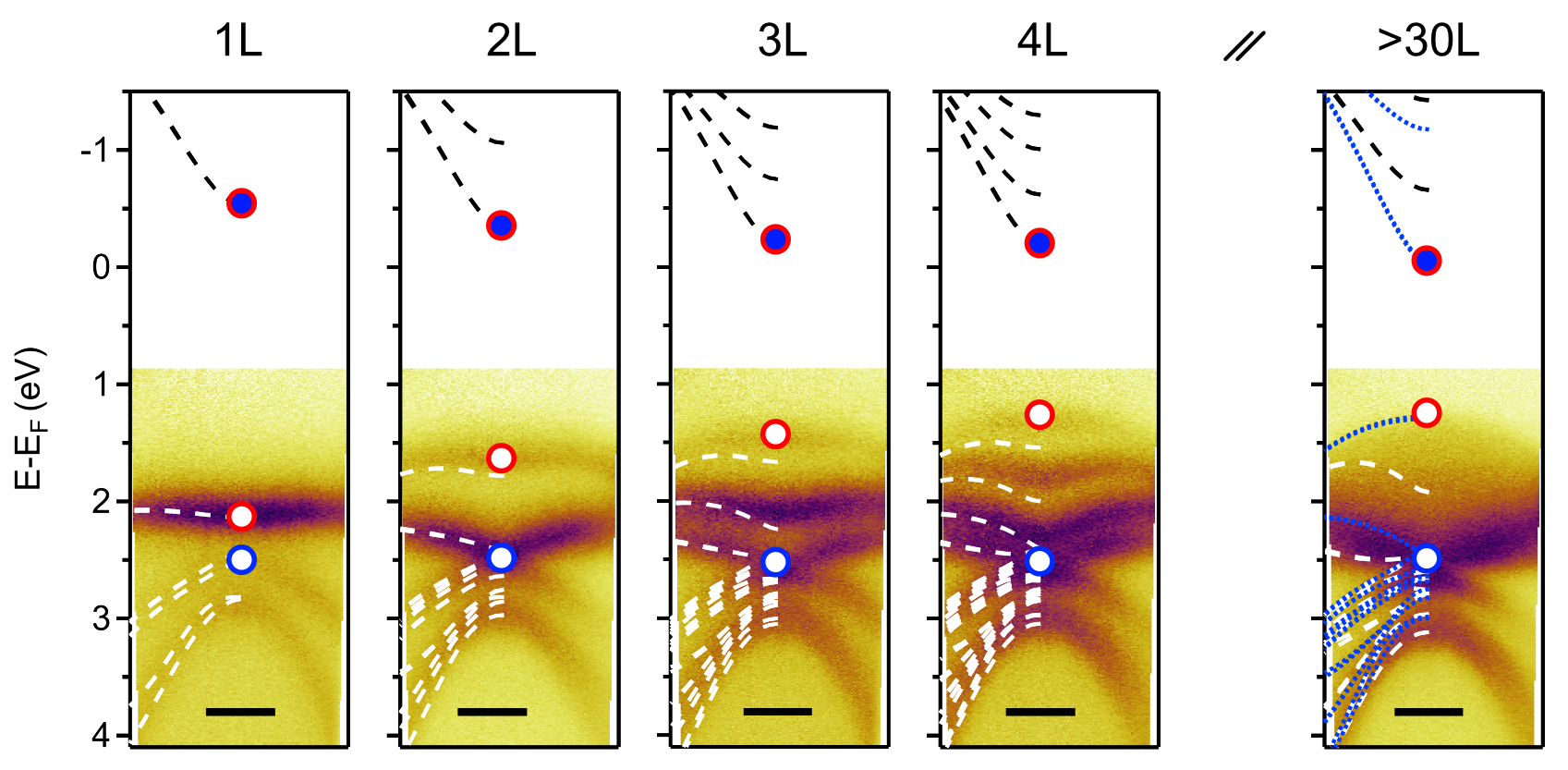}
 \caption{\label{VBlayers} \textbf{Layer-dependent dispersion of InSe around $\Gamma$.}
Energy momentum slices, $\mathrm{I}\!\left(E,\mathbf{k_\parallel}\right)$, about $\mathbf{\Gamma}$
for 1L, 2L, 3L, 4L and bulk InSe as marked. Overlaid as white (black) dashed lines are DFT predictions
for the corresponding valence (conduction) band dispersions. For bulk InSe, DFT band dispersions with $k_z=0.5$
(in units of out-of-plane reciprocal lattice parameter) are also overlaid as blue dotted lines.
Empty red (blue) circles mark the $v$ ($v_1^{\pm \frac32}$) band edge at $\mathbf{\Gamma}$;
filled circles mark the $c$ band edge. Scale bars 0.2\AA$^{-1}$.} 
\end{figure}

In Figure \ref{VBlayers} we show how the valence band spectrum changes as a function of the number of layers in a film.
In the centre of the Brillouin zone, the valence band $v$ is composed primarily of selenium P$_z$ orbitals \cite{Magorrian2016} which
overlap and hybridize between the consecutive layers, leading to the pronounced subband structure in the ARPES spectra (two subbands in a bilayer,
three in a trilayer, and so on). Simultaneously, the VBM moves to a progressively lower binding energy and gets closer to the $\mathbf{\Gamma}$-point.
For 2L InSe there remains a measurable band inversion, with $\Delta k$\SI{=0.1\pm0.1}{\per\angstrom} and $\Delta E_2$\SI{=30\pm20}{meV}.
However, for L$\geq 3$ the band inversion is no longer measurable with $\Delta E_{L\geq 3}$\SI{<20}{meV}, which is less than thermal
energy at room temperature. At the same time, the band gap and, consequently, A- and B-exciton energies are smaller in thicker films,
following the interlayer hybridisation of electron P$_z$ and S orbitals at the conduction $c$ and valence $v$ band edges.
By comparison, the $\mathbf{\Gamma}$-point edges of $v_1$ and $v_2$ bands remain almost unchanged,
due to the in-plane character of selenium's P$_x$ and P$_y$ orbitals from which it is composed \cite{Magorrian2016} (see  Table \ref{table2}).
As a result, the thickness dependence is more pronounced for A-excitons than for B-excitons,
with the difference between these two PL lines showing the same dependence on $L$ as $v-v^{\pm \frac32}_1$
measured using ARPES, (see Table \ref{table1}).

\vspace{0.5cm}

\begin{table}
\begin{tabular}{c|c c c c | c c}
  \hline
  % after \\: \hline \cline{col1-col2} or \cline{col3-col4} ...
   & $-v_1^{\pm \frac32}$ (eV) & $-v$ (eV) & $X_B$ (eV) & $X_A$ (eV) & $v-v_1^{\pm \frac32}$ (eV) & $X_B-X_A$ (eV)\\
   \hline
  1L & $2.4 \pm 0.1$ & $2.08 \pm 0.02$ & $2.96 \pm 0.05$ & $2.84 \pm 0.05$& $0.3 \pm 0.1$ & $0.12 \pm 0.07$ \\
  2L & $2.50 \pm 0.05$ & $1.63 \pm 0.02$ &  $2.60 \pm 0.10$  & $1.90 \pm 0.10$ & $0.87 \pm 0.05$ &  $0.71 \pm 0.13$ \\
  3L & $2.54 \pm 0.05$ & $1.47 \pm 0.02$ &  $2.60 \pm 0.09$  & $1.67 \pm 0.08$ & $1.07 \pm 0.05$ &  $0.93 \pm 0.12$  \\
  4L & $2.50 \pm 0.05$ & $1.32 \pm 0.02$ &  $2.58 \pm 0.11$  & $1.62 \pm 0.07$ & $1.18 \pm 0.05$ & $0.96 \pm 0.13$  \\
  5L & $2.52 \pm 0.05$ & $1.32 \pm 0.02$ &  $2.54 \pm 0.11$  & $1.47 \pm 0.07$ & $1.20 \pm 0.05$ & $1.08 \pm 0.13$  \\
  7L & $2.52 \pm 0.05$ & $1.27 \pm 0.02$ &  $2.48 \pm 0.08$  & $1.39 \pm 0.05$ & $1.25 \pm 0.05$ & $1.10 \pm 0.10$ \\
  bulk& $2.46 \pm 0.05$ & $1.30 \pm 0.02$ & $2.43 \pm 0.10$  & $1.26 \pm 0.04$ & $1.16 \pm 0.05$ & $1.17 \pm 0.10$  \\
  \hline

\end{tabular}
\caption{\label{table2} \textbf{Band parameters as a function of layer number in the film.}
Energies of the upper valence band ($v$) and lower valence band ($v_1$) at $\mathbf{\Gamma}$ measured by ARPES and
compared to A ($X_A$) and B ($X_B$) exciton energies in InSe films with various number of layers.}
\end{table}

\vspace{0.5cm}

\begin{figure}[t]
 \centering
 \includegraphics[width=8cm]{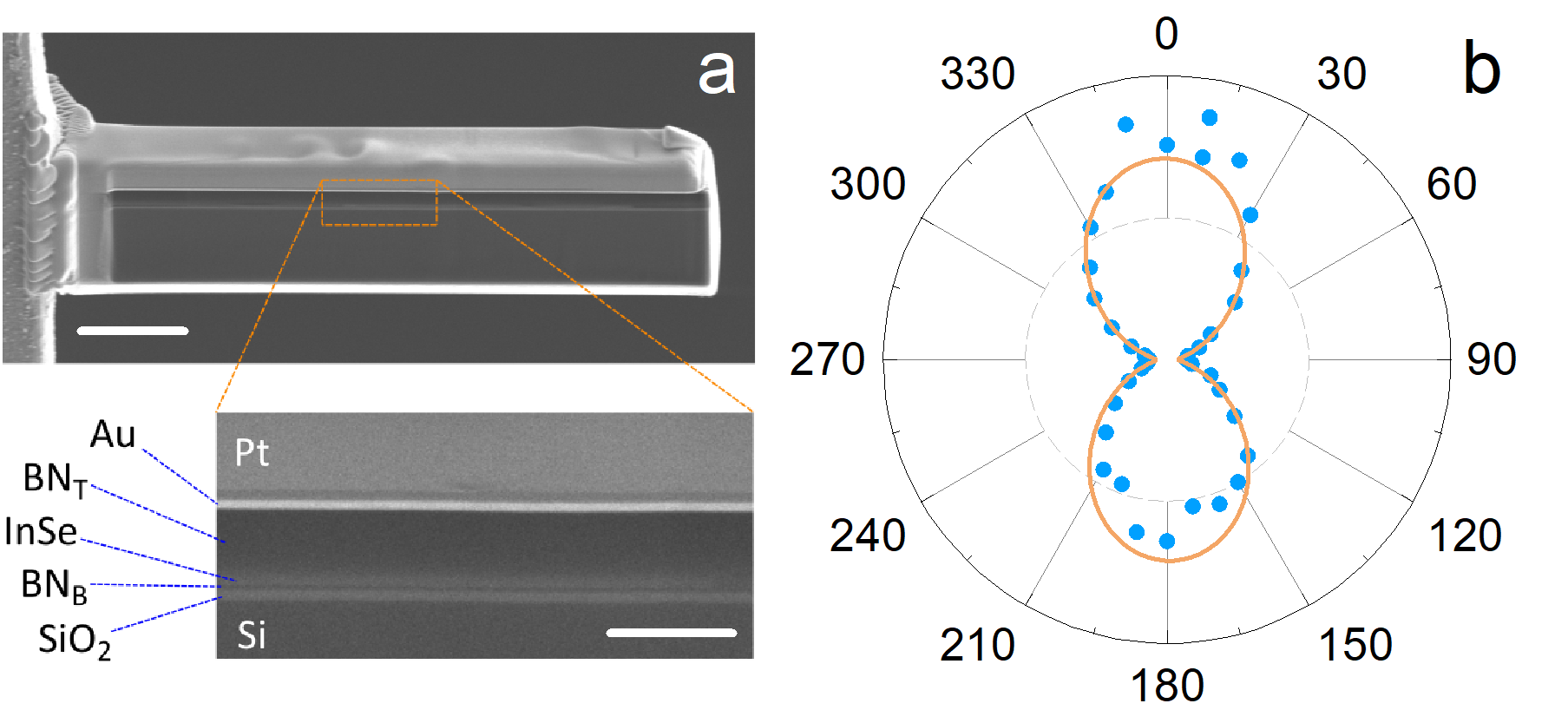}
 \caption{\label{PL_pol} \textbf{Polarized PL emission from few-layer InSe.} (a) Scanning electron micrographs of a lamella extracted from hBN/InSe/hBN heterostructure using focused ion beam, attached to a TEM grid. In the higher magnification view (bottom) the layers of the lamella can be resolved, including the 9$L$ InSe flake. Scale bars are $\SI{5}{\um}$ in the upper, and $\SI{1}{\um}$ in the lower micrograph. The lamella depth is $\sim\SI{1}{\um}$. (b) Integrated intensity of A-line photoluminescence (linear scale) as a function of detected polarization angle, excited with unpolarised light. The angles match the orientation of the SEM image in (a), with \ang{90} and \ang{270} corresponding to the basal plane of the InSe layer. }
\end{figure}

Related to their orbital composition, the valence band states involved in the monolayer A- and B-excitons have different symmetries.
The monolayer lattice exhibits $z \rightarrow -z$ mirror symmetry, and bands are either odd or even with respect to this symmetry:
$v$ and $v_2$ bands are even, while $c$ and $v_1$ are odd \cite{Magorrian2016}. As a result, A-exciton recombination (transition
$c \rightarrow v$) is coupled by electric dipole moment, $d_z$, to the photons polarised perpendicularly to the plane of the 2D crystal,
emitting light along the plane of the InSe film. By contrast, P$_x$ and P$_y$ orbitals of the mirror symmetric (even) $v_1$ band provide strong
coupling of B-excitons with in-plane circularly polarised photons, emitting those in the direction perpendicular to the plane of the 2D crystal.
These selection rules for the interband transitions also persist in the few-layer InSe films. To demonstrate the above-described
anisotropy, we developed a modification of the conventional fabrication technique used to prepare
lamellae for cross-sectional scanning transmission electron microscopy (STEM) imaging \cite{Haigh2012}, whereby a block is extracted
from a hBN/InSe/hBN heterostructure using focused ion beam (see Methods) and then rotated
by \ang{90} to allow conventional optical spectroscopy in the in-plane direction of the InSe flake,
Fig. \ref{PL_pol}a. The measured angular dependence of polarization of the emitted light, shown in Figure \ref{PL_pol}b,
demonstrates strong out-of-plane linear polarisation of the A-exciton PL line.

\subsection{Conclusions}
In conclusion, we used spatially resolved photoemission spectroscopy to demonstrate that InSe undergoes a crossover
from being an indirect gap semiconductor in monolayer and bilayer 2D crystals to direct gap in thicker films. In addition, we have measured the splitting between various valence band edges in monolayers and few-layer films.
The subbands in few-layer InSe films, determined using ARPES, agree well with the trend of a fast reduction of the band gap
as the number of layers in the film increases (from \SI{\approx 2.9}{eV} in monolayer to \SI{1.25}{eV} in bulk InSe), as observed in
 photoluminescence measurements. The polarisation of the interband transition corresponding to the A-exciton confirms
that the recombination of the latter emits light along the basal plane of InSe, thus, opening up a range of optoelectronic
applications of InSe films. For instance, electrical or optical pumping of InSe, deposited as a surface coating on a waveguide,
would emit light along its axis and the area of such a coating can be easily scaled up to increase emission intensity (unlike other
semiconducting 2DM which have to be stacked vertically to increase the brightness\cite{Withers2015}) offering an opportunity
to create fibre-optics-based light-emitting diodes and other light sources.

\section{Methods}

\textit{Theory.}
We used first principles density functional theory (DFT) as implemented in the VASP code \cite{VASP} to calculate the band structure of few-layer InSe.
The lattice parameters were taken from experiments\cite{Mudd2015}: the in-plane lattice constant was set to \SI{4.00}{\angstrom} and the inter-layer spacing to \SI{8.32}{\angstrom}.
The atomic positions in the monolayer were relaxed until forces on the atoms were below \SI{0.005}{eV/\angstrom}. The plane-wave cutoff energy was \SI{600}{eV}.
We sampled the Brillouin zone with a $24\times24\times1$ grid for few-layer InSe, and with $12\times12\times12$ for bulk.
Spin-orbit coupling was taken into account during the band structure calculations. We used the local density approximation for the exchange-correlation,
which is known to underestimate band gaps hence we implemented a scissor correction by shifting the conduction band up by a constant amount such that
the calculated gap in the bulk limit becomes \SI{1.32}{eV}, the estimated bulk gap at \SI{\sim 100}{K} based on the temperature dependence of the bulk gap in
experiments\cite{Camassel1978,Millot2010,Mudd2015}.

\textit{Sample fabrication.} For $\mu$ARPES, Bridgman-grown bulk rhombohedral $\gamma$-InSe crystals were mechanically exfoliated onto silicon oxide and thin multi-terraced flakes (1$L$ to 10$L$) selected. These flakes were picked up with monolayer graphene using the poly(methyl methacrylate) (PMMA) dry peel transfer technique \cite{Ponomarenko2011} stamped onto a laterally large (\SI{>50}{\um}), thin (\SI{<50}{\nm}) graphite (Figure \ref{mexican} and \ref{VBlayers}) or hBN (Figure \ref{Ek}) crystal on a Ti/Pt (\SI{3}{\nm} / \SI{20}{\nm}) coated highly n-doped Si wafer. The stack was annealed at \SI{150}{\celsius} for 1 hour, to induce contamination trapped between the flakes to agglomerate through the self-cleaning mechanism \cite{Haigh2012}. The exfoliations, transfers and annealing all took place within an Ar glovebox. The thick flake analysed in Figure \ref{Ek}d was only partially encapsulated by graphene which is why no graphene bands are observed in the $\mu$ARPES spectra.

For investigation of in-plane PL, InSe was exfoliated onto silicon oxide in an Ar glovebox. Suitably thin flakes were selected by their optical contrast, picked up with thin hBN (\SI{<10}{\nm}) using the PMMA dry peel transfer technique\cite{Ponomarenko2011}, and stamped onto thick hBN (\SI{>100}{\nm}) on silicon oxide. The sample was removed from the glovebox, out-of-plane PL checked, an additional layer of thick (\SI{>100}{nm}) hBN transferred on top of the encapsulated InSe, and then coated with \SI{3}{\nm} AuPd + \SI{5}{\nm} amorphous carbon for added protection. In an FEI Helios focused ion beam (FIB) dualbeam scanning electron microscope, a thick platinum (\SI{\sim 1}{\um}) layer was deposited on the 2D stack to act as an etch mask while a lamella was cut. FIB milling was used to remove a cross-section, cutting around a lamella through the stack using a \SI{30}{kV} Ga$^+$ beam, current \SI{7}{\nA} to dig trenches, then current \SI{1}{\nA} to thin the lamella. A micromanipulator was used to extract the lamella, rotate it by \ang{90}, and position it on an OMICRON transmission electron microscopy (TEM) grid. After milling, the damaged edges of the 2D stack were removed by FIB polishing using decreasing acceleration voltages (\SI{5}{ kV}, \SI{47}{\pA} and \SI{2}{kV}, \SI{24}{\pA}). The final thickness of the specimen was \SI{\sim1}{\um}.

\textit{$\mu$ARPES.} Spectra were acquired at the Spectromicroscopy beamline of the Elettra light source \cite{Dudin2010}. A Schwarzchild mirror focused \SI{27}{eV} linearly polarized radiation to a submicrometer diameter spot, striking the sample at an angle of incidence of \ang{45}. Photoemitted electrons were collected by a hemispherical analyser with 2D detector which can be rotated relative to the sample normal, aligned with the sample at its eucentric position, with energy and momentum resolution of \SI{\sim 50}{meV} and \SI{\sim 0.03}{\per\angstrom}. Samples were annealed at up to \SI{650}{K} for $> 2$ hours in ultrahigh vacuum before analysis. Liquid nitrogen cooling during measurement gave a nominal sample temperature of \SI{100}{K}. The kinetic energy of photoemitted electrons at the chemical potential, $E_F$, was determined locally on the samples by fitting a Fermi function to the drop in intensity of the graphene bands at $E_F$. The crystallographic orientation of the flakes was determined by acquiring three-dimensional energy-momentum maps, $\mathrm{I}\!\left(E,\mathbf{k_\parallel} \right)$, of the upper valence bands around $\mathbf{\Gamma}$ and identifying the high-symmetry directions. Energy-momentum slices through the Brillouin zone in high-symmetry directions were acquired by interpolation from a series of closely-spaced detector slices along the relevant directions. The energy of the upper valence band for the 1$L$ data, as shown in Figure \ref{mexican}, was determined by fitting a Lorentzian function to energy distribution curves, $\mathrm{I}\!\left(E\right)$, at each $\left(k_x,k_y\right)$ value.

\textit{Photoluminescence spectroscopy.} Monolayer PL was recorded under \SI{3.8}{\eV} illumination at room temperature in a Horiba LabRAM HR Evolution system with beam spot size of \SI{\sim2}{\um}, laser power of \SI{1.2}{mW}, and grating of $600$ grooves per millimetre. For other thicknesses, an excitation energy of \SI{2.33}{eV} was used. The polarization dependence in Figure \ref{PL_pol} was recorded at \SI{4}{K} under \SI{2.4}{eV} and \SI{0.4}{mW} unpolarised excitation with a linearly polarized detection. Each point is the result of integration of the PL signal between \SI{1.23}{eV} and \SI{1.49}{eV} for the given polarization angle.

%IN SAMPLE FABRICATION SECTION ...To prepare the lamellae, InSe specimen has been encapsulated in thick hBN (~1 µm) and coated with $3$ nm AuPd $+ 5$ nm amorphous carbon $+ 1\mu$m Pt to provide protection during 30 kV Ga+ milling. The final thickness of the specimen was ~1 micron after polishing with 16 kV, then 5 kV.

\begin{acknowledgement}
We acknowledge support from EPSRC grants EP/N509565/1, EP/P01139X/1, EP/N010345/1 and EP/L01548X/1 along with the CDT Graphene-NOWNANO, and the EPSRC Doctoral Prize Fellowship. In addition, we acknowledge support from the  European Graphene Flagship Project, ERC Synergy Grarant Hetero2D,the ARCHER National UK Supercomputer RAP Project e547, Royal Society URF, and Llyod Register Foundation Nanotechnology grant.
All data presented in this paper are available at .... to be inserted upon acceptance.... Additional data related to this paper may be requested from the authors.
\end{acknowledgement}

%\bibliographystyle{unsrt}
%\bibliographystyle{plainnat}
%\bibliography{InSeARPESv20}

\providecommand{\latin}[1]{#1}
\providecommand*\mcitethebibliography{\thebibliography}
\csname @ifundefined\endcsname{endmcitethebibliography}
  {\let\endmcitethebibliography\endthebibliography}{}

\end{document}